\documentclass[fleqn,usenatbib]{mnras}
\usepackage[dvipsnames]{xcolor}
\usepackage{newtxtext,newtxmath}
\usepackage[normalem]{ulem}
\usepackage{comment}

\usepackage[T1]{fontenc}
\usepackage{ae,aecompl}
\usepackage{times}
\usepackage{graphicx}	
\usepackage{amsmath}	
\usepackage{subfigure}

\newcommand{\btoa}{b/a}
\newcommand{\ctoa}{c/a}

\def\R500c{R_{\rm 500c}}
\def\R200m{R_{\rm 200m}}
\def\M500c{M_{\rm 500c}}

\newcommand{\lcdm}{\Lambda{\rm CDM}}

\definecolor{hpurple}{HTML}{7E16DF}

\title[Shapes \& Formation History of DM haloes]{Correlations between Triaxial Shapes and Formation History of Dark Matter haloes}

\author[Lau et al.]{Erwin  T.\ Lau
,$^{1}$\thanks{E-mail: erwin.lau@miami.edu}
Andrew P. Hearin$^{2}$, 
Daisuke Nagai$^{3,4}$,
Nico Cappelluti$^{1}$
\\
$^1${Department of Physics, University of Miami, Coral Gables, FL 33124, U.S.A.}\\
$^2${Argonne National Lab, Lemont, IL 60439, U.S.A.}\\
$^3${Department of Physics, Yale University, New Haven, CT 06520, U.S.A.}\\
$^4${Department of Astronomy, Yale University, New Haven, CT 06520, U.S.A.}\\
}

\pubyear{2020}

\begin{document}
\label{firstpage}
\pagerange{\pageref{firstpage}--\pageref{lastpage}}
\maketitle

\begin{abstract}
The shape of dark matter haloes plays a critical role in constraining cosmology with upcoming large-scale structure surveys. In this paper, we study the correlations between the triaxial shapes and formation histories in dark matter haloes in the MultiDark Planck 2 $N$-body cosmological simulation. We find that halo ellipticity is strongly correlated with halo properties that serve as proxies of halo formation history, such as halo concentration and the normalized peak-centroid offset. These correlations are nearly independent of the halo density peak height. We present a simple model for the correlation between halo ellipticity and concentration using conditional abundance matching, and provide fitting formulae for the multi-dimensional distributions of triaxial halo shape as a function of halo peak height. We apply our halo shape model to gauge the effects of halo ellipticity and orientation bias on the excess surface mass density profiles in cluster-size haloes. Our model should be useful for exploring the impact of triaxial halo shape on cosmological constraints in upcoming weak lensing surveys of galaxy clusters. 
\end{abstract}


\section{Introduction}

In the $\lcdm$ cosmological model, the fundamental building block of large-scale structure is the dark matter (DM) halo. The centers of haloes are the deepest points in the gravitational potential of the Universe, and so DM haloes of sufficient mass are natural sites of galaxy formation \citep{white_rees78, blumenthal_etal84}. Observations of cluster-mass haloes contain rich information about cosmology \citep[e.g.,][]{allen_etal11}, and so a long-standing goal of large-scale structure cosmology is to accurately model and characterize the abundance, spatial distribution, and internal structure of DM haloes. 

The shape of DM haloes is generically expected to be non-spherical due to the non-spherical shape of the Gaussian density peak of the primordial density field from which the halo forms \citep{Doroshkevich1970} and the directional nature of the merger and accretion of haloes along filamentary structures in the cosmic web \citep{zeldovich70}. Since DM halo shape depends upon both the initial density field and on halo assembly, the observed distribution of halo shapes can be used to test and validate the $\lcdm$ structure formation scenario \citep{kawahara10, sereno_etal18}. The shape of galaxy clusters can also serve as a probe of the fundamental particle nature of DM: models of self-interacting DM generically predict more spherical distributions of cluster mass relative to models in which DM is collisionless, an effect that becomes more pronounced in the inner regions \citep{yoshida_etal2000,spergel_steinhardt2000, dave_etal01,peter_etal13}.

The non-spherical shape of DM haloes has important implications for the study of cosmology and astrophysics with observations of galaxy clusters. Specifically, the shapes of DM haloes affect the measurements of the mass and gas content in galaxy clusters, which are commonly assumed to be spherically symmetric. In particular, the non-spherical shape of haloes can lead to biases gravitational lensing estimates of cluster mass due to the elongation of the mass distribution along the line-of-sight \citep[e.g.][]{meneghetti_etal10,lee_etal18}.
This leads to what is known as orientation bias \citep{hanama_etal12,dietrich_etal14,Osato_2018}. With orientation bias, elliptical clusters elongated along the line-of-sight are preferentially detected, with their masses over-estimated.
Hydrostatic estimates of cluster masses can also be under- or over-estimated as a result of the failing of the spherical symmetry assumption \citep[e.g.][]{chiu_molnar12,buote_humphrey12b}. The non-spherical shape of haloes is therefore a source of systematic uncertainty in cosmological constraints derived from a wide variety of measurements of galaxy clusters \citep[e.g.,][]{smith_watts05, battaglia_etal12b}. 

Previous theoretical studies have already established the triaxial nature of DM haloes \citep[e.g.,][]{Jing_2002}, where DM halo shape can be well-approximated by a triaxial ellipsoid specified by two parameters: the ratios of minor- to major- and the intermediate- to major axes. The distribution of halo shapes exhibits a clear dependence on halo mass, with higher mass haloes being more elliptical relative to haloes of lower mass; at fixed mass, DM haloes at higher redshift present more elliptical shapes relative to present-day haloes \citep{kasun_evrard05, Allgood_2006}. The mass-dependence of the average halo shape is approximately universal \citep{bonamigo_etal15, Vega_Ferrero_2017}, but there is significant scatter in halo shape at fixed halo mass. The scatter of halo shape at fixed mass can largely be attributed to differences in the halo formation histories \citep{chen_etal19}, as halo shapes are known to correlate with the halo age \citep[e.g.][]{despali_etal14, despali_etal17}, and to evolve with time as the halo grows \citep{Suto_2016}. The dependence of shapes of haloes on their assembly histories are reflected in their correlations with their local environments \citep{jeeson-daniel_etal11, morinaga_ishiyama20, chen_etal20}. 

In this paper, we study the correlation between DM halo shape and other halo properties that are readily measured from cosmological simulations. In particular, we focus on how such correlations manifest in the gravitational lensing signal of group- and cluster-mass haloes. 
We build a simple analytical model that captures how the two-dimensional distribution of halo shapes varies across redshift with halo mass, halo concentration, and various halo formation proxies. We use our model to quantify how halo ellipticity and triaxiality contributes to scatter in the lensing of haloes of fixed mass, and we give a proof-of-principle demonstration that our model has the capability to derive constraints on the distributions of halo shapes across redshift using stacked lensing measurements of galaxy clusters. In addition, we provide a simple model for the surface mass density of DM halo that depends on halo mass, concentration, ellipticity, and orientation, which will be useful for assessing systematics in weak lensing measurements in upcoming optical cluster surveys.

This paper is organized as follows. In \S\ref{sec:shape} we give an overview of how we quantify DM halo shape, and in \S\ref{sec:sims} we describe the simulations we used to validate our model. We present our model for the distribution of triaxial halo shapes in \S\ref{sec:results}, and in \S\ref{sec:cam} we extend this model to incorporate correlations with halo formation history and concentration. We apply our shape model to assess the effects of halo shape and orientation on cluster lensing signals in \S\ref{sec:orientation-bias}. In \S\ref{sec:discussion} we discuss our results in the context of previous work, and conclude in \S\ref{sec:conclusion} with a summary of our principal findings. 

\section{Definitions of Halo Shape}
\label{sec:shape}

For an ellipsoidal halo with major, intermediate, and minor axes of length $a, b,$ and $c,$ the shape of the halo can be described in terms of two parameters, {\em ellipticity} and {\em prolaticity}, defined as follows:
\begin{eqnarray}
\label{eq:ellipticity}
{e} &\equiv& \left(1 - (\ctoa)^2\right)/2{L} \\
{p} &\equiv& \left(1 - 2(\btoa)^2 + (\ctoa)^2\right)/2{L},
\end{eqnarray}
where $L \equiv 1 + (\btoa)^2 + (\ctoa)^2.$
We furthermore define halo {\em triaxiality}, $T,$ as:
\begin{eqnarray}
T &\equiv& \frac{1-(\btoa)^2}{1-(\ctoa)^2} = \frac{1}{2}\left(1+\frac{p}{e}\right). 
\end{eqnarray}
The condition $a\geq b \geq c\geq0$ implies that the domain of $e$ is the range $[0, 1/2]$, and $T$ is the range $[0,1],$ while the domain of $p$ is $[-e,e]$. Note that there is a physical lower limit of $T \geq 2-1/(2e)$, otherwise $\ctoa$ will not be a real number. 
Haloes with $T \gtrsim 2/3$ are called {\em prolate} and present an elongated, cigar-like shape, while {\em oblate} haloes with $T \lesssim 1/3$ exhibit a flattened shape like a lentil or a disk. In the present work, we will build our model for halo shape using $e$ and $T$ as independent variables, but we note that the equations above make it straightforward to transform these quantities to other alternative shape variables that are in common usage. 

\section{Simulation}
\label{sec:sims}

In this paper, we aim to study the effects of halo formation histories on halo shapes for group- and cluster-size haloes. For this purpose, we analyze the MultiDark Planck 2 (MDPL2) cosmological simulation  \citep{klypin_etal16} that contains a large number of highly-resolved group- and cluster-size haloes with well measured formation histories. The MDPL2 is a gravity-only $N$-body simulation of $3840^3$ particles in a periodic box with $L_{\rm box}=1{\rm Gpc}/h,$ giving a particle mass resolution of $m_{\rm p} \approx 1.51 \times 10^9 h^{-1}M_\odot$. The MDPL2 was run with a  flat cosmology similar to \citet{planck2013}, with $h=0.6777$, $\Omega_{\rm m} = 0.307115$, $\Omega_\Lambda = 0.692885$, $\sigma_8 = 0.829$ and $n_{\rm s} = 0.96$. We refer the reader to \citet{klypin_etal16} for more details about the simulation. 
Throughout this work, we use the axis ratio measurements and proxies for halo assembly histories from the publicly available ROCKSTAR \citep{rockstar,consistent_trees,bpl_rockstar} halo catalogs in the MDPL2 simulation; data products for MDPL2 are publicly available through the MultiDark Database \citep{riebe_etal13}, and can downloaded from the {\tt CosmoSim} website.\footnote{\url{https://www.cosmosim.org}}

We select distinct host haloes\footnote{That is, ROCKSTAR subhaloes for which {\tt upid}=-1.} with virial mass $M_{\rm vir} \geq 10^{13} h^{-1} M_\odot$ at $z=0.0, 0.5, 1.0, 1.5$ haloes in MDPL2 in this mass range are resolved by at least $2000$ particles within the virial radius, $R_{\rm vir}$, such that their shape measurements are robust.   

We use the axis-ratio measurements and halo formation parameters provided in the ROCKSTAR halo catalog. The axis ratios are computed following the iterative method outlined in \citet{Allgood_2006}. For each halo, the substructure-excluded mass tensor for DM particles is computed as $$I_{ij} \equiv \frac{1}{N}\sum^N_p x_i x_j,$$ starting with all particles within the virial radius. Here $x_i$ is the position of the particle in the direction $i = 1,2,3$ relative to the halo centre. The major, intermediate, and minor axes $(a,b,c)$ of the ellipsoid are then the square roots of the sorted eigenvalues of the mass tensor. In the iteration steps, every particle with ellipsoidal radius $r_{\rm ep} = a\sqrt{(x/a)^2 + (y/b)^2+(z/c)^2} \leq R_{\rm vir}$ (where $x,y,z$ are the coordinates of the particle in the frame of eigenvectors) is included in the computation of the mass tensor. The iteration repeats until the axis ratios converge to within a given tolerance. 

We examine the correlation of halo ellipticities with 10 commonly used halo formation proxies:
\begin{enumerate}
    \item Halo concentration $c_{\rm vir} \equiv R_{\rm vir}/R_s$, where $R_s$ is the scale radius found by fitting the DM density profile to the Navarro, Frenk and White (NFW) profile \citep{nfw96};
    \item Virial ratio $T/|U|$, computed as the ratio of the kinetic energy of the DM particles within the virial radius to the absolute value of gravitational potential energy of same DM particles;
    \item Half-mass scale $a_{1/2}$, which is scale factor at which the halo attains half of its present mass found by tracing the main progentior line of the halo;
    \item The time difference in Gyr between now and when the halo reaches a half of its current mass $\Delta t_{1/2}$;
    \item The difference in collapse density $\Delta \omega_{1/2} = \omega(z_{1/2})-\omega(z_0)$ between now and when halo reaches a half of its current mass, where $\omega(z) \equiv 1.686/D(z)$ and $D(z)$ is the linear growth factor; 
    \item The scale factor at which the halo experiences its last major merger $a_{\rm lmm}$;
    \item Peak-centroid offset $X_{\rm off}$, computed as the distance between the density peak and center of mass scaled by the virial radius;
    \item Peak-centroid Velocity Offset $V_{\rm off}$, computed as the magnitude difference in mean halo velocity and velocity of the densest peak, $V_{\rm off}$;
    \item The instantaneous mass accretion rate $\dot M_{\rm vir}$, computed as the virial mass change between two consecutuve snapshots;
    \item The change in virial mass over the past dynamical time $\Delta M_{\rm vir}/\Delta T_{\rm dyn}$.
\end{enumerate}
These proxies and their physical meanings are summarized in Table~\ref{tab:X}. 
We refer the reader to \citet{rockstar,bpl_rockstar} for further details on how these halo properties are measured. 

\begin{table*}
\begin{center}
\begin{tabular}{|c|l|}
\hline
Parameter & Physical meaning\\
\hline
$c_{\rm vir}$ & Halo concentration \\
$T/|U|$ & Virial Ratio = ratio of kinetic to potential energy of the halo = 0.5 for a completely virialized halo \\
$a_{1/2}$ & Half-mass scale, which is the scale factor at which halo reaches a half of its current mass. \\
$\Delta t_{1/2}$ & time difference in Gyr between now and when halo reaches a half of its current mass. \\
$\Delta \omega_{1/2}$ & Difference in re-scaled `time' $\omega \equiv \delta_c(z)$ between now and when halo reaches a half of its current mass. \\
$a_{\rm lmm}$ & Scale factor at which halo experiences its last major merger. \\
$X_{\rm off}$ & Distance offset between mass centroid and densest peak normalized by the virial radius. \\
$V_{\rm off}$ & Velocity offset between mean halo velocity and velocity of the densest peak. \\
$\dot M_{\rm vir}$ & Instantaneous mass accretion rate. \\
$\Delta M_{\rm vir}/\Delta T_{\rm dyn}$ & Mass change over the past dynamical time. \\
\hline
\end{tabular}
\end{center}
\caption{\label{tab:X} Table of halo formation proxy parameters.}
\end{table*}

\section{Results}
\label{sec:results}

\subsection{Distributions of Ellipticity and Triaxiality}
\label{subsec:ellipticity_triaxility_one_points}

\begin{figure}
\begin{center}
\includegraphics[width=0.95\linewidth]{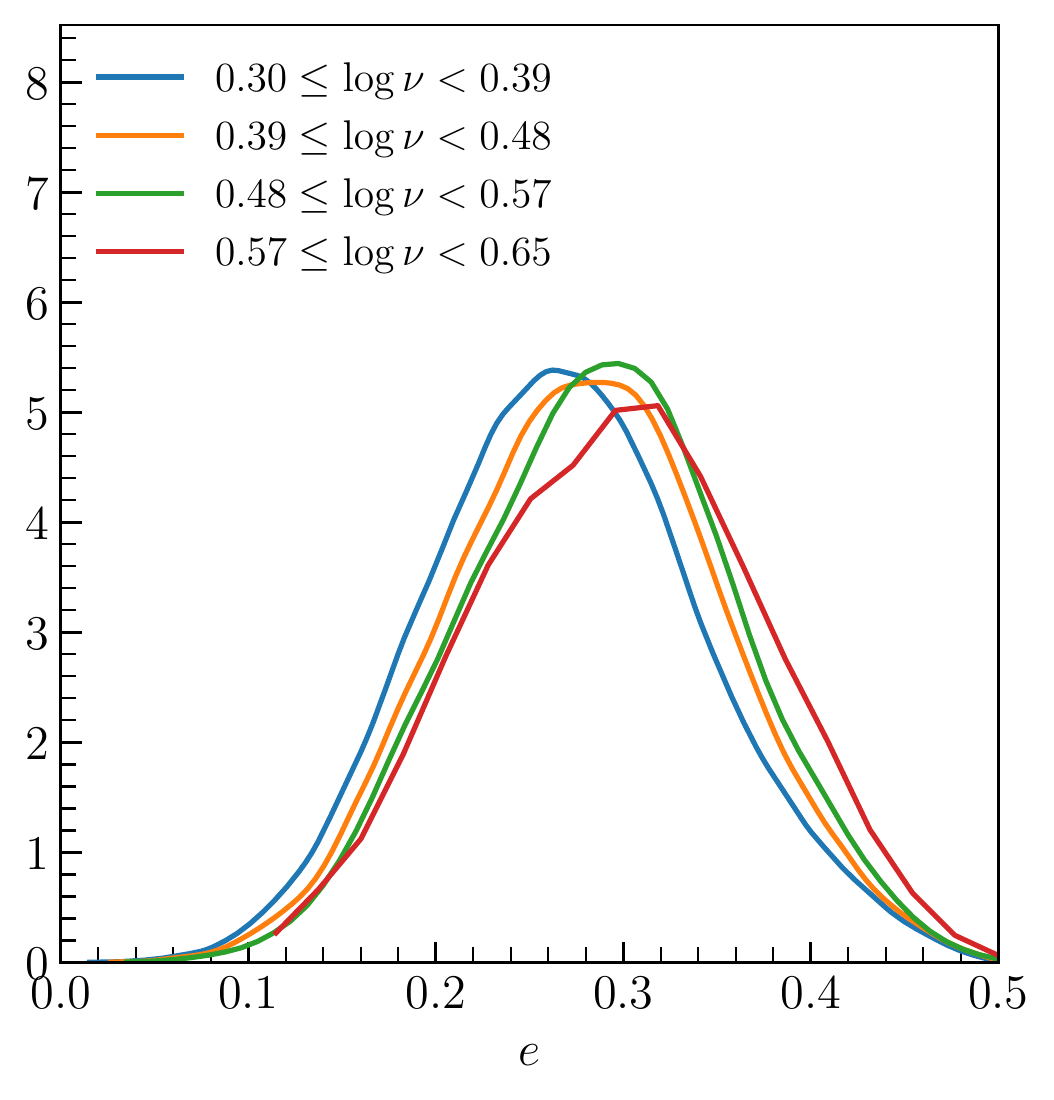}
\caption{Normalized probability density distribution of ellipticity in bins of peak height (in $\log_{10}$) for haloes with $M_{\rm vir} \geq 10^{13} h^{-1}M_\odot $ for all haloes at $z=0, 0.5, 1.0, 1.5$. While haloes with higher peak height have slightly higher average ellipticities, the distributions broadly overlap at all mass.}
\label{fig:edist_zall}
\end{center}
\end{figure}

To characterize the joint mass- and redshift-dependence of halo ellipticity, we use the halo density peak height, $\nu\equiv1.686/\sigma(M_{\rm vir},z),$ to represent the mass of the halo at a given redshift, where $\sigma(M_{\rm vir},z)$ is the mass density fluctuation of a halo of mass $M_{\rm vir}$ at redshift $z$. 
Figure~\ref{fig:edist_zall} shows the distributions of ellipticity $e$ at different peak height values (in $\log_{10}$) for $M_{\rm vir} \geq 10^{13} h^{-1}M_\odot$ at $z=0, 0.5, 1.0, 1.5$ haloes with larger peak height (i.e., more massive and lower redshift haloes) tend to have slightly higher mean ellipticities, however there is quite significant overlap in the full distribution of shapes of haloes in different mass bins. The shape for the highest $\nu$ bin is slightly different from the rest due to smaller number of haloes contained in the bin.  At any particular value of $\nu,$ the distribution of halo ellipticity is well-described by the generalized gamma distribution
\begin{eqnarray}\label{eq:ellip_dist}
P(e | \zeta, \eta) &=& \frac{|\zeta|e^{\zeta\eta-1}\exp(-e^\eta)}{\Gamma(\zeta)}, 
\end{eqnarray}
where $\Gamma(z) = \int_0^\infty x^{z-1}e^{-x}dx$ is the gamma function, and $\zeta$ and $\eta$ are two free parameters that govern the shape of the distribution. We thereby capture the mass- and redshift-dependence of the ellipticity distribution via the following calibration:
\begin{eqnarray}
\zeta &=& 1.0 + 1.7(\log \nu / 0.25), \\
\eta &=& 3.1.
\end{eqnarray}

\begin{figure}
\begin{center}
\includegraphics[width=1.0\linewidth]{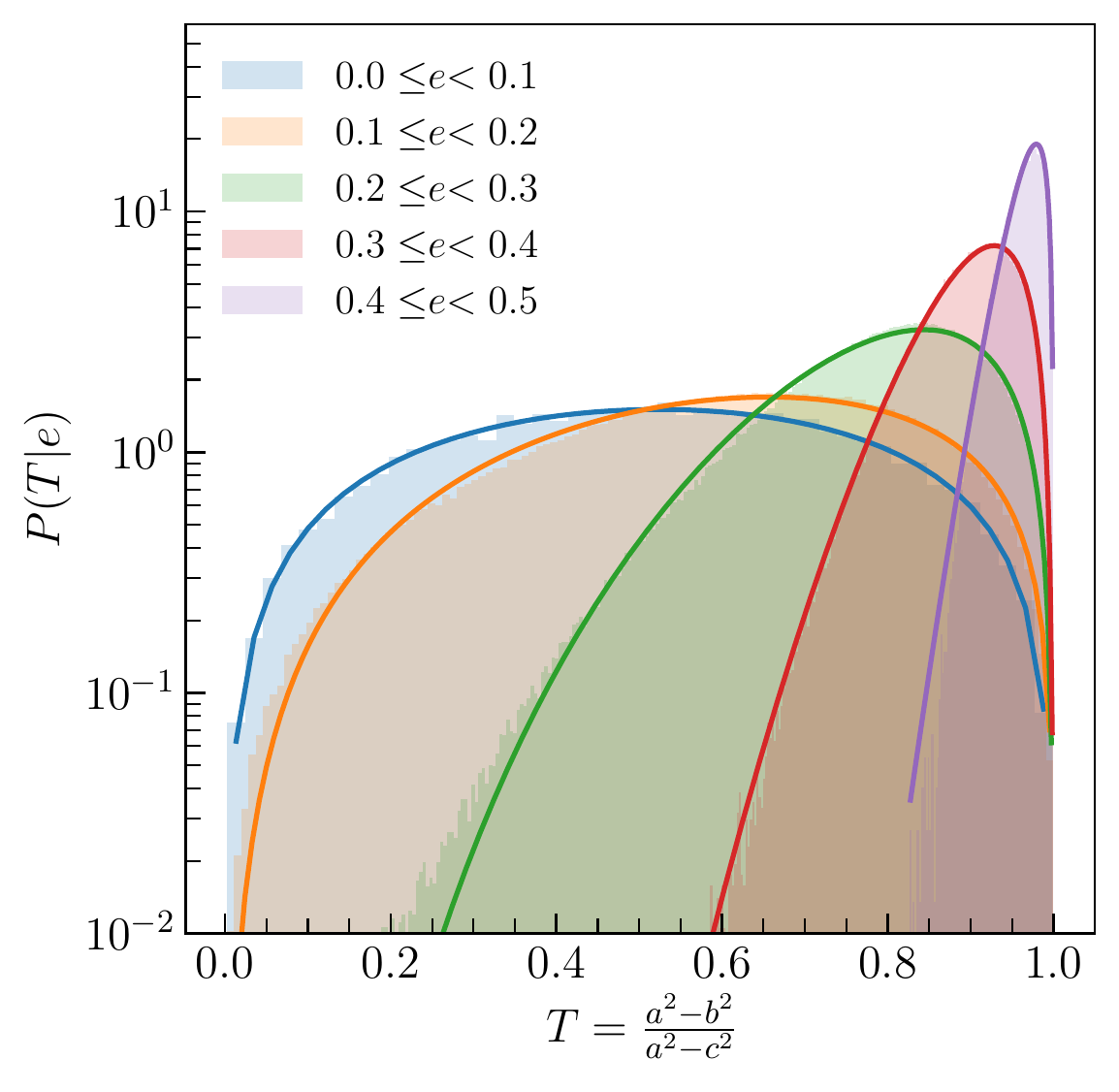}
\caption{Normalized probability density distribution of triaxiality $T$ for all haloes at $z=0,0.5,1.0$ in bins of ellipticity $e$ as indicated in the legend. Solid lines show beta distributions that have been fit to the measured ellipticity distributions (see Eq.~\ref{eq:triax_dist}).}
\label{fig:tdist_z0}
\end{center}
\end{figure}

In addition to ellipticity, we also need to specify triaxiality to completely describe the shape of an ellipsoidal halo. Figure~\ref{fig:tdist_z0} shows the distributions of triaxiality at fixed ellipticity, $P(T|e),$ for all haloes at $z=0,0.5,1.0,1.5$. haloes with larger values of $e$ also tend to have larger values of $T,$ meaning that highly elliptical haloes tend to be preferentially prolate. The spread of the triaxiality distribution also depends on halo ellipticity, such that less elliptical haloes tend to have a broader distribution of triaxiality. 

For a population of haloes with ellipticity $e,$ we model $P(T|e),$ the conditional distribution of $T,$ with the beta distribution:
\begin{eqnarray}\label{eq:triax_dist}
P(T| e; \xi, \chi) = \frac{\Gamma(\xi+\chi)}{\Gamma(\xi)\Gamma(\chi)}T^{\xi}(1-T)^{\chi},
\end{eqnarray}
where $\Gamma(z)$ is the gamma function, and $\xi, \chi > 0$ are two free parameters that govern the shape of the distribution. 
We find that the beta distribution parameters $(\xi, \chi)$ depend on halo ellipticity $e$ in a manner that is well-described by the following calibration:
\begin{eqnarray}
    \xi &=& \chi\left(\frac{1-3.5e+25e^2}{1-2e}\right),\\
    \chi &=& 2.5.
\end{eqnarray}

\subsection{Correlation between halo ellipticity and formation history proxies}
\label{subsec:ellipticity_zform_correlations}

\begin{figure}
\begin{center}
\includegraphics[width=1.0\linewidth]{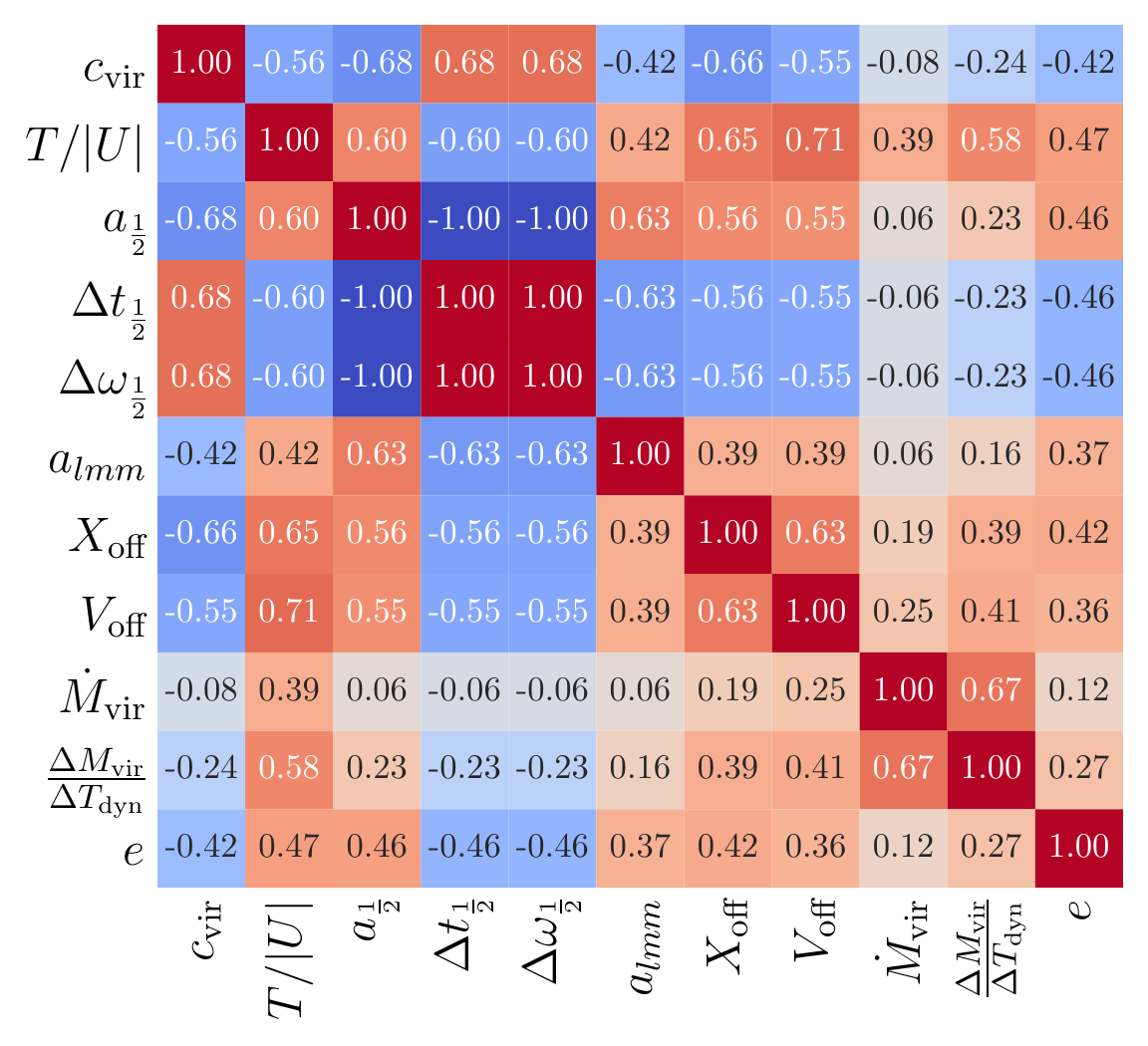}
\caption{Matrix of the Spearman rank correlation coefficients $\rho_{\rm sp}$ computed among halo ellipticity $e$ and different halo mass accretion proxies in Table~\ref{tab:X}, for haloes at  $z=0$.  } 
\label{fig:corr_mat_z0}
\end{center}
\end{figure}

\begin{figure}
\begin{center}
\includegraphics[width=1.0\linewidth]{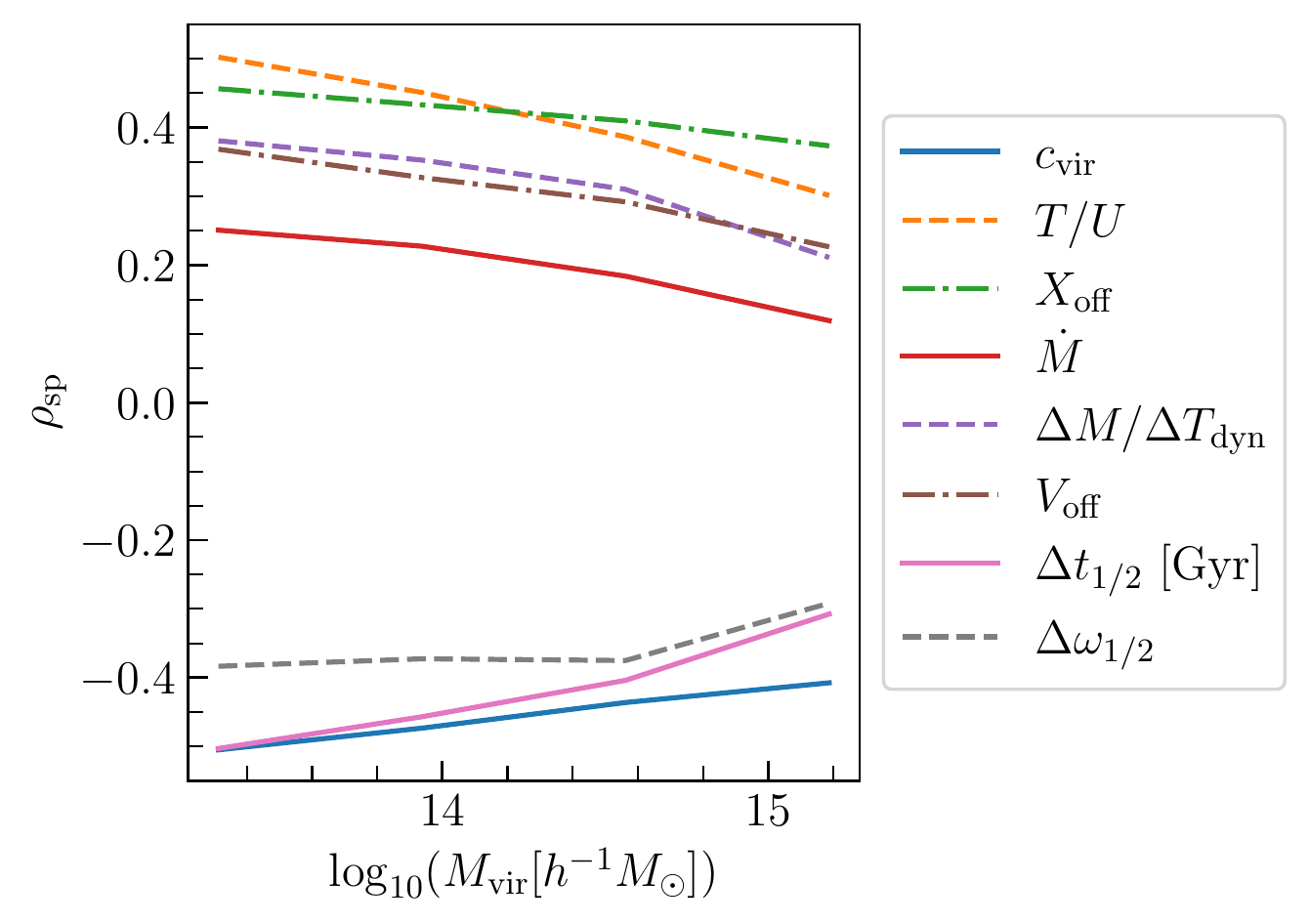}
\caption{Dependence of the Spearman rank correlation coefficient $\rho_{\rm sp}$ between halo ellipticity $e$ and different halo mass accretion proxies (shown by different coloured lines), plotted as a function of halo mass for haloes at $z=0, 0.5, 1.0, 1.5$. Halo concentration $c_{\rm vir}$ correlates (or anti-correlates) most strongly with halo ellipticity, followed closely by $T/|U|$, $\Delta t_{1/2}$, and $X_{\rm off}$ (see Table~\ref{tab:X} for details). All the correlation coefficient shows weak mass dependence. Note that we exclude the correlation coefficient for $a_{1/2}$ and $a_{lmm}$, since they measure the halo growth differently for haloes at different redshifts. } 
\label{fig:corr_mat_mdep}
\end{center}
\end{figure}

In this section we examine the correlation between halo ellipticity $e$ and the halo formation proxies listed in Table~\ref{tab:X}. We compute the Spearman rank correlation coefficient $$\rho_{\rm sp}(p_i,p_j) = \mathrm{Cov}(p_i,p_j)/(\sigma_i\sigma_j)$$ between two rank-ordered halo parameters $p_i, p_j$, where $\sigma_i$ is the standard deviation for parameter $p_i$, and $\mathrm{Cov}(p_i,p_j)$ is the covariance between $p_i$ and $p_j$. Note that all of reported correlations are statistically significant, as they all have $p$-values much less than $0.05$.

Figure~\ref{fig:corr_mat_z0} shows the matrix of the correlation coefficients computed for haloes at $z=0$. The correlation between $e$ and the virial ratio $T/|U|$ is the tightest, with $\rho_{\rm sp} = 0.47$. This is closely followed by $a_{1/2}$, with $\rho_{\rm sp} = 0.46$. The correlation coefficient between $e$ and the other two formation time proxies $\Delta t_{1/2}$ and $\Delta \omega_{1/2}$ are the same as that for $a_{1/2}$, with their signs flipped. This is because at $z=0$, $a_{1/2}$, $\Delta t_{1/2}$, and $\Delta \omega_{1/2}$ only differ by multiplicative constants. The next most correlated quantities are the halo concentration $c_{\rm vir}$ with $\rho_{\rm sp} = -0.42$, and  the normalized density peak offset $X_{\rm off}$ with $\rho_{\rm sp} = 0.42$. The scale factor of the last major merger $a_{lmm}$, velocity offset $V_{\rm off}$, instantaneous mass accretion rate $\dot{M}_{\rm vir}$ and mass change of the dynamical time $\Delta M_{\rm vir}/\Delta T_{\rm dyn}$, all show less correlation with $e$ compared other parameters.  Note that the halo formation proxies presented here also correlate strongly with each other. For example, the halo concentration $c_{\rm vir}$ is shown to anti-correlate strongly with the halo formation time $a_{1/2}$, with $\rho = -0.68$. This is evident from the strong anti-correlation between $c_{\rm vir}$ and $e$, and the strong correlation between $a_{1/2}$ and $e$. 
These results are consistent with previous works studying halo shape correlations \citep[e.g.,][]{maccio_etal07,skibba_maccio11,wang_mo_etal11, jeeson-daniel_etal11}.

When we consider all haloes at all four redshifts $z=0.0, 0.5, 1.0, 1.5$, the halo concentration is most correlated with ellipticity, with $\rho_{\rm sp} = -0.50$ , closely followed by the virial ratio and the half-time difference $\Delta t_{1/2}$ with $\rho_{\rm sp} = 0.49$, and the density peak offset $X_{\rm off}$ with $\rho_{\rm sp} =0.46$. These values are very close to those for haloes at $z=0$ only.  However, we notice that the correlation coefficient for the half-mass scale is drastically reduced to $\rho_{\rm sp} = 0.03$ when considering haloes at all redshifts. This is an artifact because half-mass scale is measured with respect to $z=0$ (or $a=1$), which means that it does not measure the same half-mass growth time for haloes at other redshifts. This is also true for $a_{lmm}$. The other two time measures, $\Delta t_{1/2}$ and $\Delta \omega_{1/2}$, are better proxies for comparing the half-mass growth for haloes at multiple redshifts. Figure~\ref{fig:corr_mat_mdep} shows the mass dependence of $\rho$ for all haloes at $z=0.0, 0.5, 1.0, 1.5$. The correlation coefficients shows little dependence on halo mass differently for different halo formation proxies. This is especially the case for halo concentration. The slight mass dependence in $\rho$ is due to the difference in the mass dependence of concentration and ellipticity.  The concentration increases and the ellipticity decreases for lower halo mass. The increase in concentration is faster than the decrease in ellipticity, leading to a slight decrease in the correlation coefficient. 
  
\begin{figure*}
\begin{center}
\includegraphics[width=1.0\linewidth]{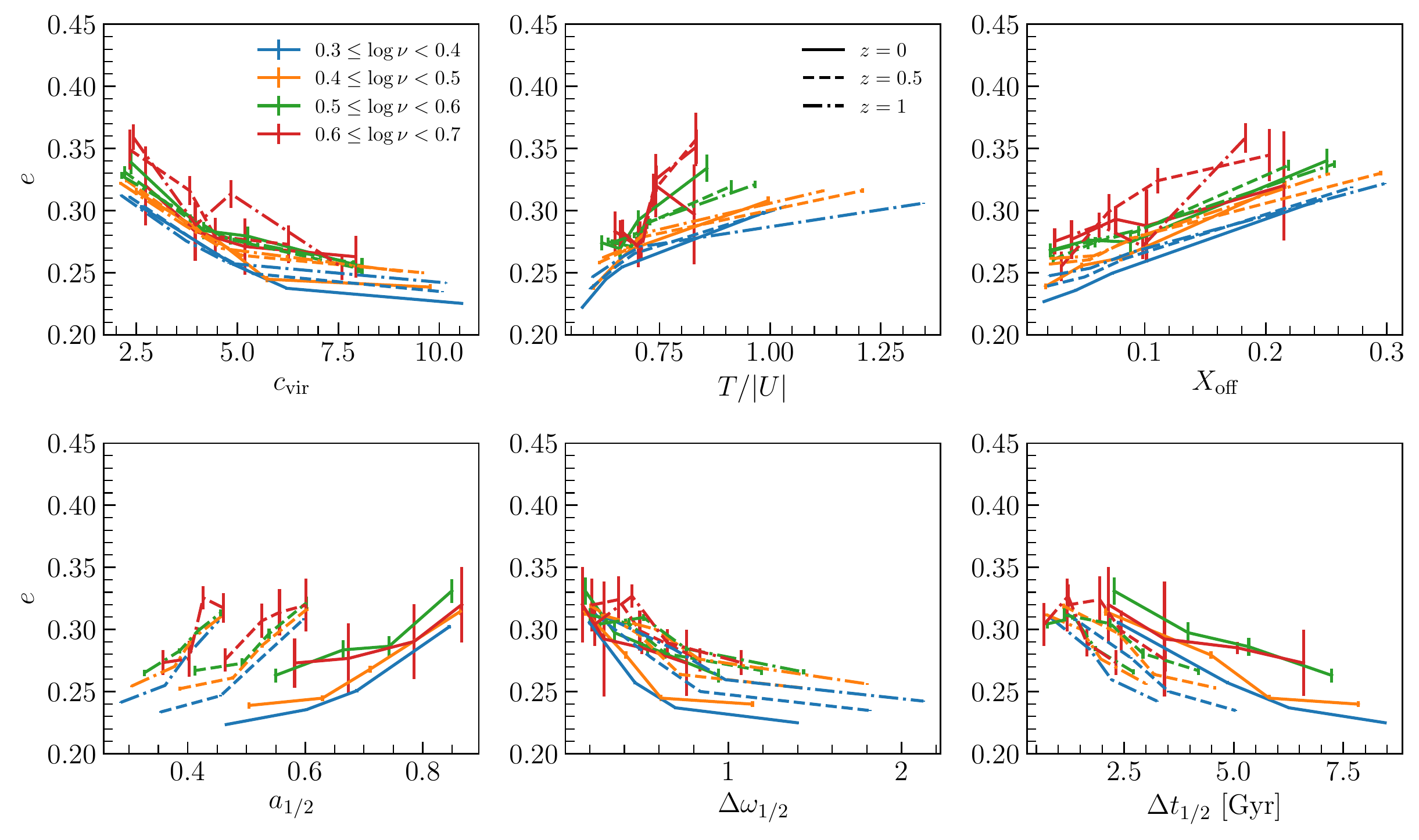}
\caption{Dependence of ellipticity on $c_{\rm vir}$, $T/|U|$, $X_{\rm off}$ (top row, left to right), $a_{1/2}$, $t_{1/2}$, and $\Delta \omega_{1/2}$ (bottom row, left to right), for different peak height (in $\log_{10}$) bins at $z=0, 0.5, 1.0$ (shown as solid, dashed, and dot-dashed lines respectively). Error bars indicate $1\sigma$ error on the mean. Correlations with $c_{\rm vir}$ and $X_{\rm off}$ show only very weak dependence upon either peak height or redshift.}
\label{fig:ellip_mass_z_all}
\end{center}
\end{figure*}

Next we examine dependence of the correlation between halo ellipticity and the halo formation proxies on halo peak height $\nu$. 
Figure~\ref{fig:ellip_mass_z_all} shows the dependence of ellipticities on the halo formation proxies for different halo peak height bins
at redshifts $z=0, 0.5, 1.0$. We drop the $z=1.5$ bin due to the relatively small number of massive haloes at that redshift. Here we only shows the correlations for the halo formation proxies that show the tightest correlations with halo ellipticity, namely, $c_{\rm vir}$, $T/|U|$, $X_{\rm off}$, $a_{1/2}$,  $\Delta t_{1/2}$ and $\Delta \omega_{1/2}$.  
In general, higher peak height haloes (i.e., more massive haloes) have higher ellipticities, which is expected since they are still actively accreting mass from the surrounding and thus most `unrelaxed' and elliptical. 

Different halo formation proxies show variations in their $\nu$ and redshift dependence. For ``time''-based proxies: $a_{1/2}$, $\Delta t_{1/2}$, and $\Delta \omega_{1/2}$, their relations with ellipticity varies the most between halo masses and redshifts. Note that these quantities differ only up to multiplicative constants that depend on cosmology and redshift, so they correlations with halo ellipticity at the same redshift are almost identical. On the other hand, ``space''-based proxies: $c_{\rm vir}$, $T/|U|$, $X_{\rm off}$, show less dependent on halo mass and redshift, as these quantities are computed at a single snapshot based on the spatial and kinematic structures of the halo, thus they are less sensitive to the background cosmology, compared to the time-based proxies. 

\section{Modeling halo ellipticity using conditional abundance matching}
\label{sec:cam}

\begin{figure*}
\begin{center}
\includegraphics[width=0.45\linewidth]{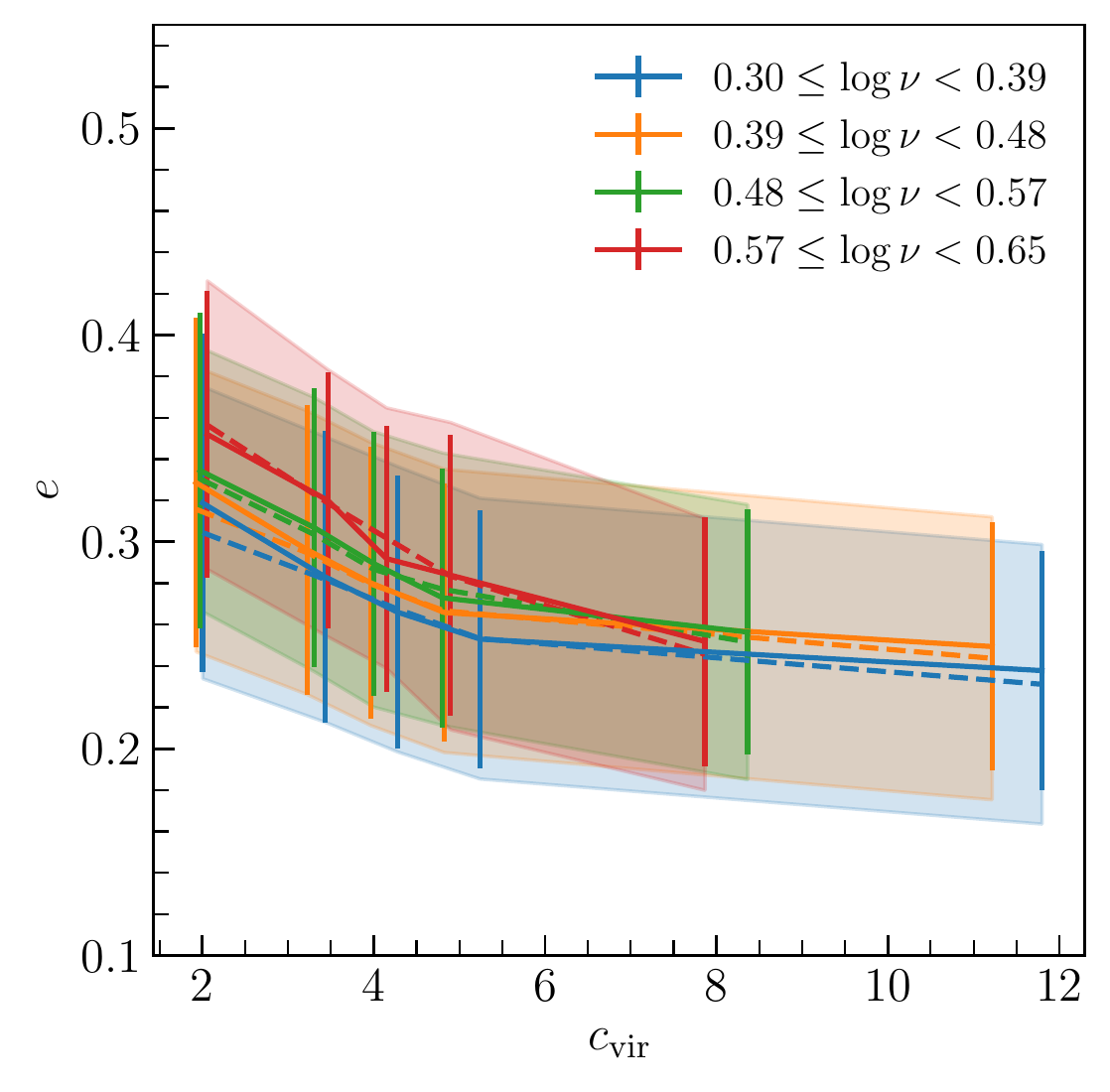}
\includegraphics[width=0.435\linewidth]{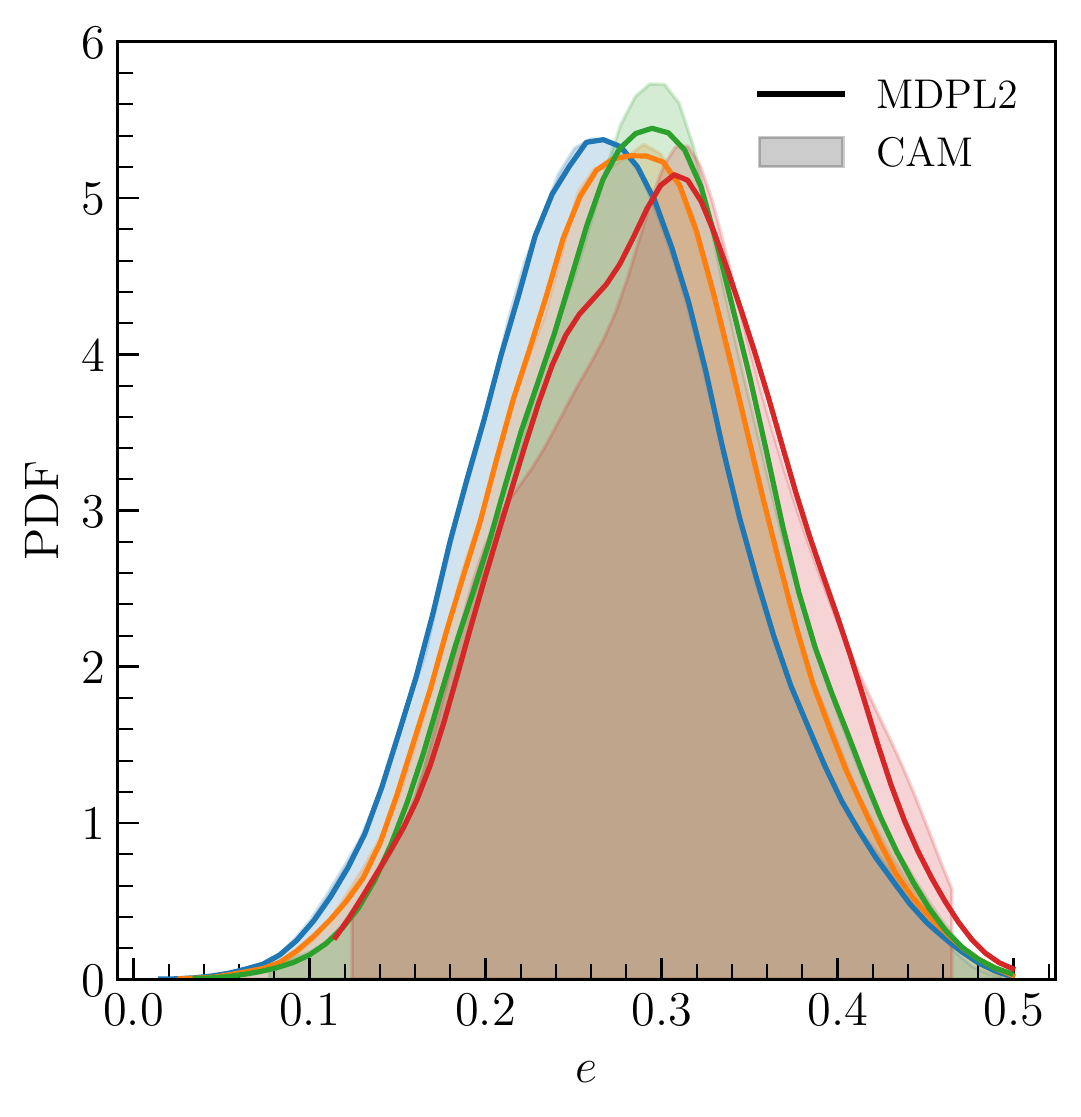}
\caption{{\em Left panel}: Dependence of ellipticity upon $c_{\rm vir},$ with error bars show $1\sigma$ variance amongst all haloes at $z=0, 0.5, 1.0, 1.5$ Dashed lines show the prediction of the probabilistic model computed using Conditional Abundance Matching (CAM) with  shaded regions indicating $1\sigma$ variance. Different colors indicate different halo peak height bins, $\log_{10} \nu,$ as indicated in the legend. {\em Right panel:} Distribution of ellipticity in the same bins of peak height, $P(e; c_{\rm vir}| \log_{10} \nu)$, directly measured from simulation (solid lines), and those in our CAM model (shaded regions).}
\label{fig:cam_model}
\end{center}
\end{figure*}

In this section we present a probabilistic model for the dependence of halo ellipticity upon halo assembly. The basis of our model is the Conditional Abundance Matching technique \citep[CAM,][]{masaki_etal13,hearin_etal13}, coupled together with the results in Section~\ref{sec:results}. While CAM has capability to additionally capture higher dimensional correlations \citep{hearin_etal20}, here we focus on capturing the dependence of ellipticity upon halo concentration at fixed halo mass, as we found $c_{\rm vir}$ be the secondary halo property with the strongest correlation with halo ellipticity. We note that the CAM formalism could similarly be applied to any other proxy for halo formation, provided that the $M_{\rm halo}$-conditioned correlation between ellipticity and the halo formation is monotonic \citep[see][section 4.3 for technical details]{hearin_etal13b}.\footnote{See the {\tt conditional\_abunmatch} function in Halotools \citep{hearin_etal17} for a publicly available implementation of CAM.}

We implement the CAM ansatz to capture the correlation between halo ellipticity and concentration by computing the cumulative distribution function (CDF) of each quantity at fixed halo peak height. The correspondence between ellipticity and concentration is then based on matching the conditional CDF of each variable. We then introduce stochasticity in the relation using the correlation coefficient computed from the simulation, resulting in a relation between ellipticity and halo formation proxy that closely follows from what we measured in the simulation, as shown in Figure~\ref{fig:cam_model}. The left panel compares the ellipticity-concentration relation between our model and the simulation, while the right panel compares the distribution of ellipticity at a given peak height. We compare the two distributions by performing a Kolmogorov-Smirnov two-sample test, finding $p$-values greater than $0.05$ at all values of peak height, so that the simulated and modeled distributions are statistically indistinguishable at the $2\sigma$ level.  

\section{Simple model of Halo Ellipticity and Orientation Bias in Weak Lensing Measurements}
\label{sec:orientation-bias}

\begin{figure*}
\begin{center}
\includegraphics[width=0.33\linewidth]{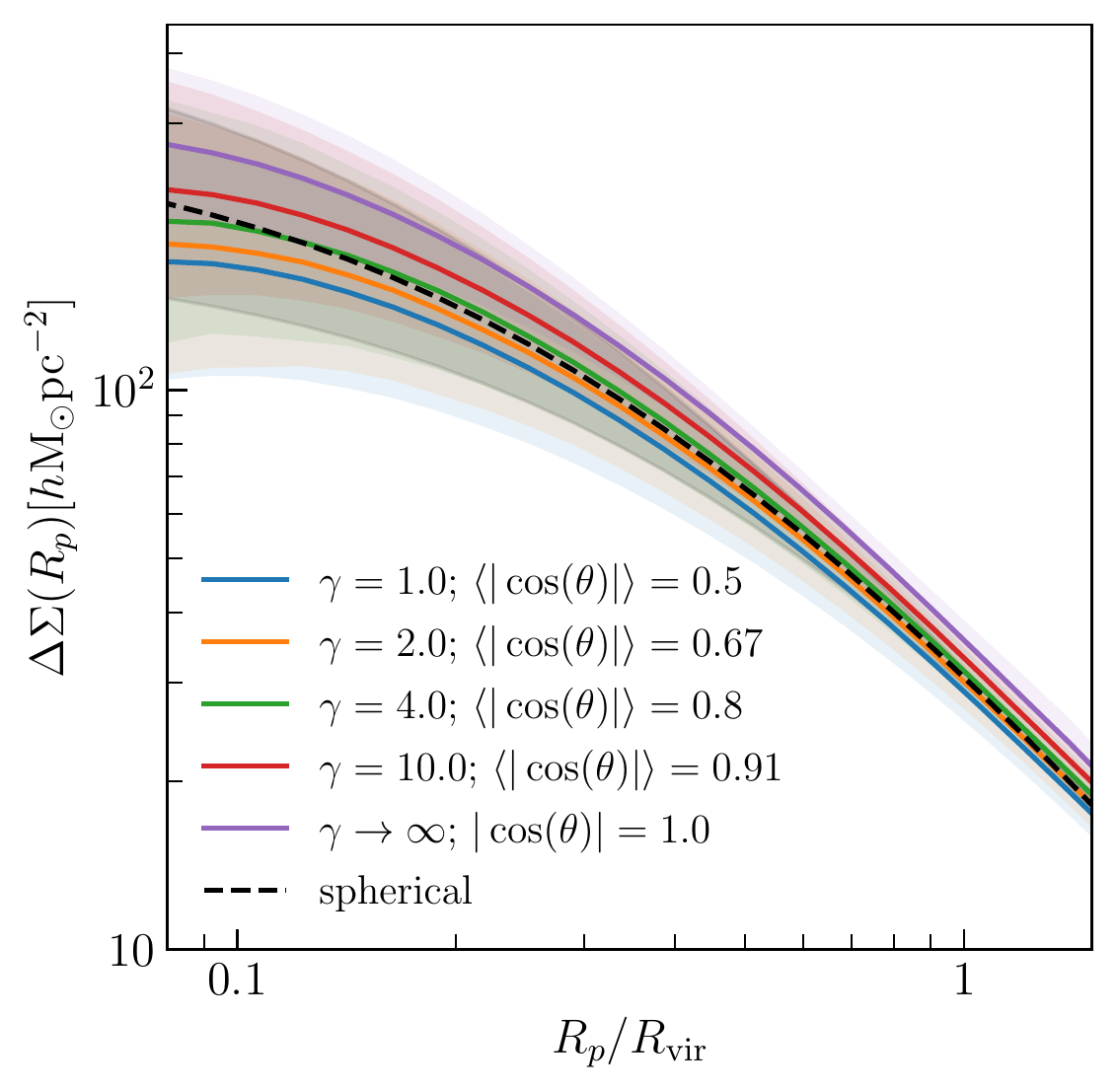}
\includegraphics[width=0.33\linewidth]{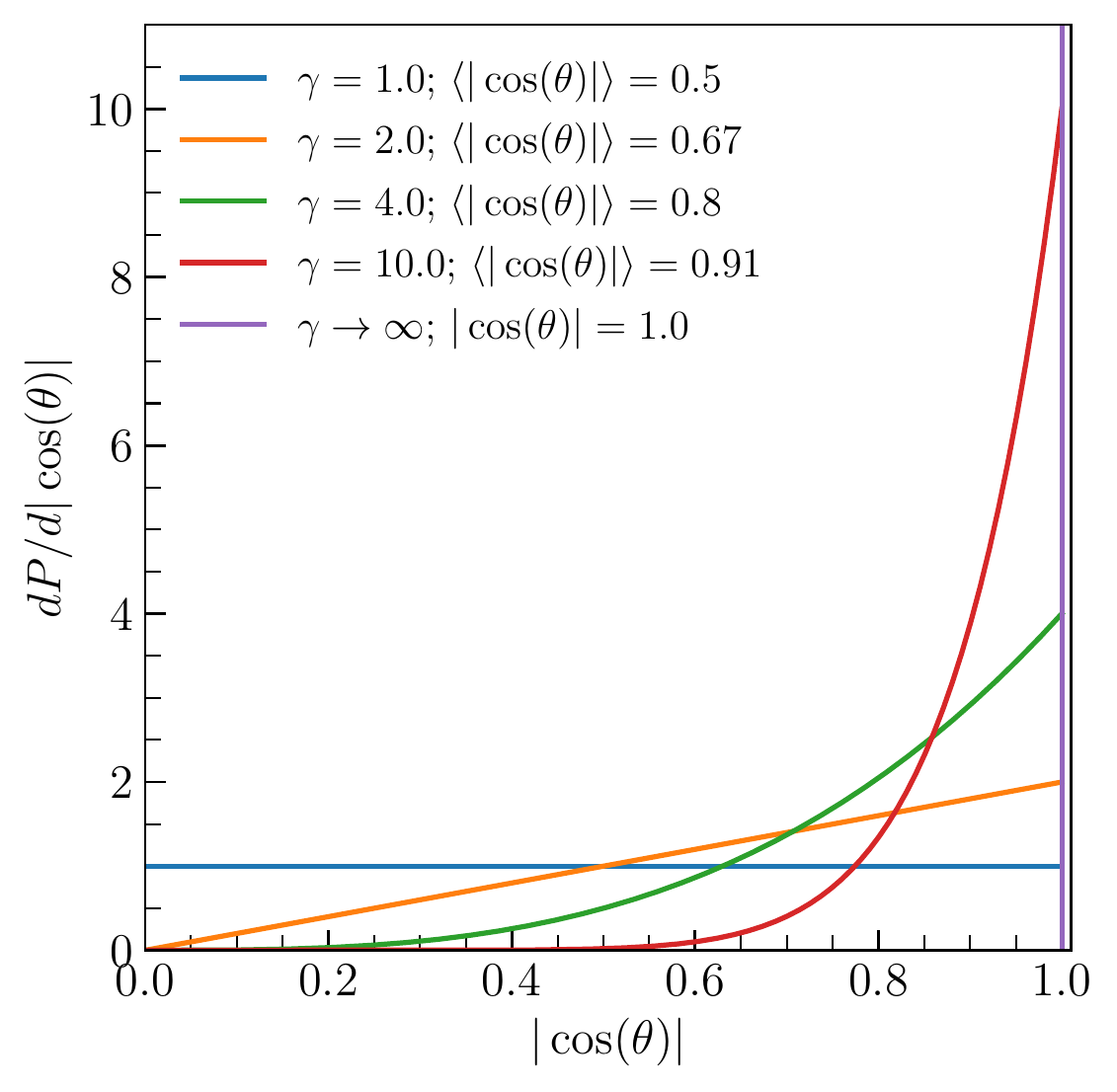}
\includegraphics[width=0.33\linewidth]{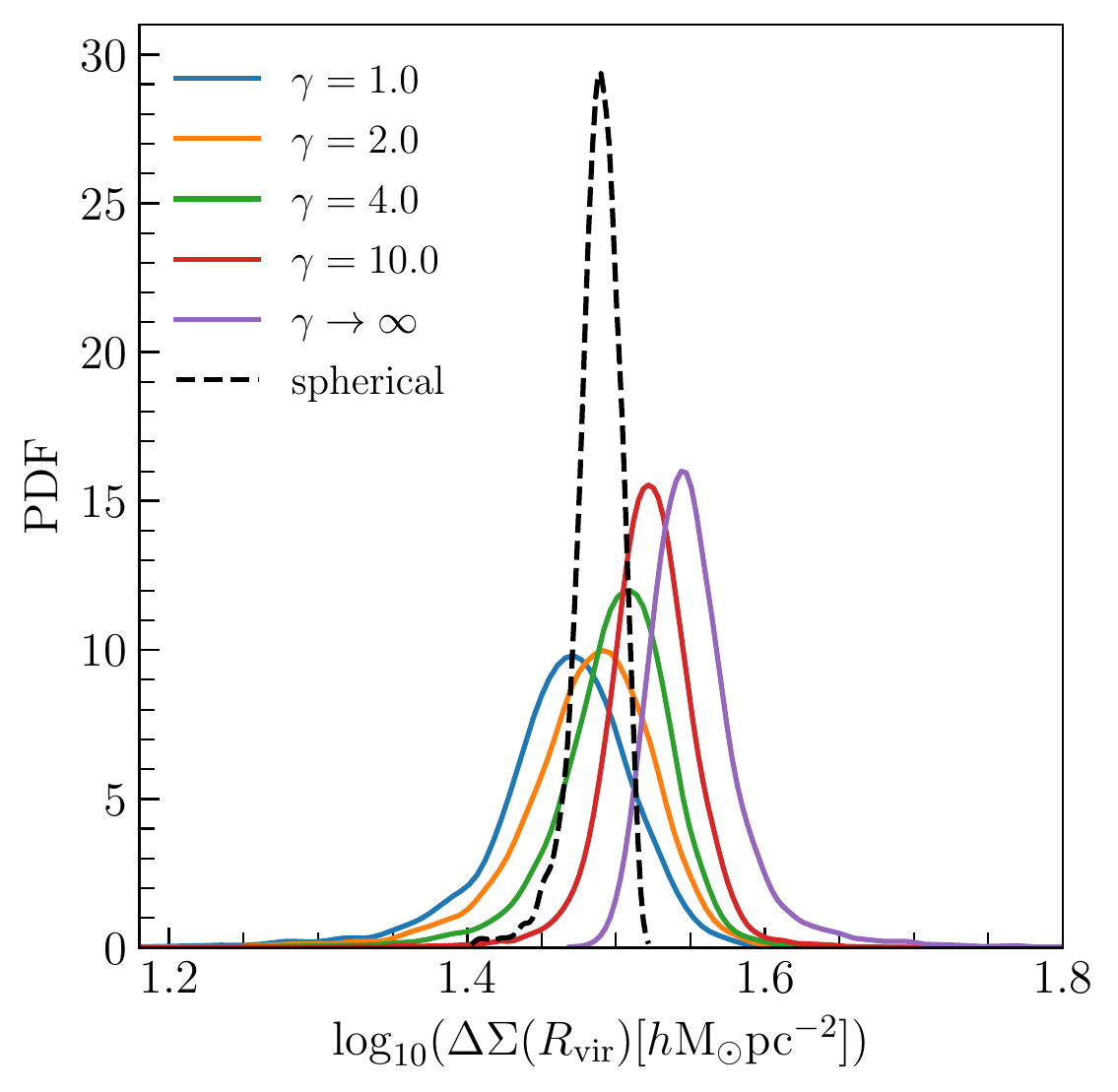}
\caption{
Impact of halo orientation bias on excess surface mass density profiles, $\Delta\Sigma,$ for haloes of mass $M_{\rm vir}\approx10^{14}M_{\odot}$ in the MDPL2 simulation at $z=0.$ {\em Left panel:} Mean $\Delta \Sigma$ profiles of triaxial haloes with different levels of orientation bias, as indicated by the parameter $\gamma$, where $\gamma = 1$ corresponds to a stack of randomly oriented haloes (i.e., zero orientation bias), and $\gamma \rightarrow \infty $ corresponds to a stack of haloes with major axes that preferentially align with the line-of-sight. The black dashed curve shows results for spherical haloes. Shaded regions indicate $1\sigma$ scatter. {\em Middle panel:} Influence of the parameter $\gamma$ on the distribution of $|\cos(\theta)|,$ with $\theta$ the angle between the principal axis and the line-of-sight. 
{\em Right panel:} Normalized probability density distribution of $\Delta \Sigma$ measured at $R_{\rm vir}$ for stacks of haloes with different values of $\gamma$. Results for spherical haloes are shown with the black dashed line.}
\label{fig:ds}
\end{center}
\end{figure*}

The distribution of the shapes of dark matter haloes, and the alignment of these shapes with large-scale structure, has a significant impact on measurements of gravitational lensing \citep{schneider_etal11}. Elliptical haloes with their major axes aligned with the line-of-sight have above-average $\Delta \Sigma$ for their mass, and are thus more efficient lenses that may be preferentially selected in observations.  Similarly, elongated low-mass clusters and groups can be mis-identified as having higher richness and higher mass, due to their more concentrated galaxy and mass distributions along the line-of-sight \citep{dietrich_etal14}. To address this issue, we present a simple model of so-called ``orientation bias'' based on the results of halo ellipticity presented in the previous sections. 

In the context of estimating cluster masses, the relevant quantity is the excess surface mass density for the 1-halo term, defined as
\begin{eqnarray}
\label{eq:dsdef}
    \Delta \Sigma (R_p) = \bar{\Sigma} (< R_p) - \Sigma(R_p), 
\end{eqnarray}
where 
\begin{eqnarray}
\label{eq:sigmaellipse}
\Sigma (R_p) &=& 2\int^\infty_{R_p}\frac{\rho dz}{\sqrt{R_p^2+z^2}}, \\
\bar{\Sigma}(<R_p) &=& \frac{1}{\pi R_p^2}\int^{R_p}_0 2\pi r dr \Sigma (r),
\label{eq:sigmaellipse2}
\end{eqnarray}
where $R_p$ is the projected radius perpendicular to the line-of-sight, $z$ is the coordinate along the line-of-sight, and $\rho$ is the 3D density profile. 

To generate our model's predictions for $\Delta \Sigma$ profiles, we adapt the Monte Carlo method used in \citet{sgro_etal13} to calculate the integrals in Equations~(\ref{eq:sigmaellipse}) and (\ref{eq:sigmaellipse2}). First, under the assumption that each halo follows the spherical NFW density profile, we use $N=10^5$ particles to sample the spherical density profile for given halo mass $M_{\rm vir}$ and concentration $c_{\rm vir}$.  We then draw ellipticity $e$ and triaxiality $T$ values from the distributions from Equations~(\ref{eq:ellip_dist}) and (\ref{eq:triax_dist}) and transform $e$ and $T$ to the axis ratios $\ctoa$ and $\btoa$. We then transform the coordinates of the particles with $\ctoa$ and $\btoa$ under the condition that the volume of the halo is conserved, i.e., $R_{\rm vir} = (abc)^{1/3}$. Assuming that $z$ is direction of the elongated axis, and the $x$ direction points to the shortest axis, we have the transformations $z \rightarrow z a^{2/3}/(bc)^{1/3}$,  $y \rightarrow y a^{2/3}/(bc)^{1/3}(b/a)$, and $x \rightarrow x a^{2/3}/(bc)^{1/3}(c/a)$. Note that in the above model, we assume NFW profiles and ignore effects of substructures, which can bias the mean and increase the scatter in the $\Delta \Sigma$ profiles. We will explore and model these effects in the future.

To model halo orientation, we assume the cosine of the angle between the elongated axis and the line-of-sight $\cos(\theta)$, follows the distribution
\begin{eqnarray}
P(|\cos(\theta)|) = \frac{\Gamma(1+\gamma)}{\Gamma(\gamma)}|\cos(\theta)|^{\gamma},
\end{eqnarray}
where $\gamma \geq 1$ is a parameter that controls the shape of the distribution, and thus controls the degree of orientation bias. When $\gamma=1$, $P(|\cos(\theta)|)$ is uniform, thus the elongated axes of the haloes are randomly oriented. Increasing $\gamma > 1$ will lead to more haloes having their elongated axes aligning with the line-of-sight. For $\gamma \rightarrow \infty$, all haloes are aligned with the line-of-sight.  
For any given value of $\gamma$, we draw a random $\cos(\theta)$ value from the distribution for each halo, and rotate the halo such that its elongated axis makes an angle $\theta$ with the line-of-sight. Finally, we compute $\Delta \Sigma$ by integrating over the particles along the line-of-sight cylinder with projected radius $R_p$. 
To show how ellipticity and halo orientation can impact lensing measurements, we use the above technique to compute $\Delta \Sigma$ for 8190 haloes with mass $M_{\rm vir} \in [0.9,1.1] \times 10^{14} M_\odot$ taken from the MDPL2 simulation at redshift $z=0$. Figure~\ref{fig:ds} shows how orientation bias impacts the lensing profiles of a stack of cluster-mass haloes. In each panel, curves of the same color correspond to a different levels of orientation bias, with the middle panel visually demonstrating the distribution of orientation angles for different values of $\gamma.$ The left panel illustrates lensing profiles of the halo samples of different orientation bias; the right panel demonstrates the diversity of lensing signals amongst haloes in the sample by showing the distribution of $\Delta\Sigma(R_{\rm vir})$ amongst members of the stack.

The dashed black curve in each panel shows results for the case of spherical haloes with the same halo mass and concentration distributions as the triaxial haloes. The dashed black and solid blue curves exactly overlap in the middle panel (i.e. they are not subject to any orientation bias), while their mean lensing profiles are discernible in the left panel, reflecting the nontrivial change to the lensing profile induced by halo ellipticity. 

Unfortunately, there is a degeneracy between the strength of orientation bias, $\gamma,$ and the true underlying distribution of ellipticity. We demonstrate this degeneracy with the green curve corresponding to $\gamma=4,$ which has an orientation bias of $\langle\vert\cos(\theta)\vert\rangle=0.8;$ from the left-hand panel we see that this sample of haloes presents a nearly indistinguishable mean lensing signal relative to a comparable stack of spherical haloes \citep[see also][for closely related discussion on cluster lensing degeneracies]{meneghetti_etal07,dietrich_etal14}. 
However, the model presented in this work is a {\em global} model for the {\em distribution} of halo shapes across mass and redshift, and by comparing the black and orange curves in the right-hand panel of Figure~\ref{fig:ds}, we can see that the scatter in the lensing profiles is tighter for a stack of spherical haloes relative to an elliptical stack with orientation bias. As discussed in \S\ref{sec:discussion}, this suggests it may be possible to extract additional information about triaxial halo profiles by incorporating our model for halo shape distributions into prediction pipelines for multi-wavelength synthetic lightcones. 
The degeneracy between ellipticity and orientation can also be broken with independent constraints on 3D triaxial shape, which can be estimated with multiwavelength datasets that include both weak and strong lensing, as well as X-ray photometry and  spectroscopy, and SZ measurements of the cluster \citep{sereno_etal18, umetsu_etal18}. 

\section{Discussion}
\label{sec:discussion}

We have provided a calibrated approximation to the full probability distribution of halo shape as a function of halo mass, concentration and redshift, $P(e, T\vert M_{\rm halo}, z, c).$ Our model for the distribution of triaxial halo shapes has potential to extend the predictive power of otherwise conventional implementations of the halo model. Standard formulations of the halo model include ingredients to predict the abundance and internal structure of DM haloes as a function of their mass and redshift. For a halo of a given mass, concentration and redshift, our model creates the capability to make a significantly richer set of predictions for the distribution and evolution of large-scale structure. Along similar lines, simulated halo catalogs may not have halo shape information available, particularly when generated by approximate $N$-body methods \citep[e.g.,][]{monaco_etal13,izard_etal16,feng_etal16}; our model offers a straightforward way to augment such catalogs with physically realistic distributions of triaxial shapes.

We note several limitations of our analysis, which assumes that (1) halo internal structure is perfectly described by the NFW profile; (2) ellipticity and triaxiality are constant functions of halo radius; and (3) substructure effects can be neglected. Realistically, halo ellipticity is known to vary with radius and can be significantly affected by the presence of substructure. Moreover, observational systematics such as mis-centering and projection effects contribute additional scatter and bias to the lensing signal $\Delta \Sigma$ that we do not account for here.  
Additionally, the current work is based on gravity-only simulations of a single {\em Planck}-like cosmology \citep{planck2018}. However, halo shape is known to depend on both cosmology \citep{ho_etal06} and baryonic effects such as radiative cooling, star formation, and feedback from supernovae and supermassive blackholes. In particular, while baryonic cooling often leads to more spherical due to more concentration of mass in the halo centres \citep{kazantzidis_etal04,lau_etal12}, feedback can often help reduce the amount of cooling, making the haloes less spherical and more inline with the gravity-only shapes  \citep{bryan_etal13,Suto_2017,chua_etal18}. Baryonic physics also impacts the relationship between halo concentration and halo assembly \citep{duffy_etal10}, a key ingredient of our approach. 

On the one hand, these caveats imply that our current calibration will not provide a high-accuracy reproduction of the distribution of halo shapes seen in hydrodynamical simulations of different cosmology. However, simulation-based predictions have capacity to capture a range of physical effects that challenge traditional halo-model approaches, and our model of triaxial halo shape has been formulated with such purposes in mind. In ongoing follow-up work, we will characterize how baryonic physics modify the distribution of halo shapes, quantifying these changes in terms of modifications to the parameters of the model presented here. We will similarly use suites of cosmological simulations to study how changes in cosmology manifest in changes to the distribution of halo shapes. Multi-wavelength lightcone maps constructed via our forward model thereby create an opportunity to derive cluster-based constraints on cosmology in a manner that is robust to systematic uncertainty in baryonic physics by self-calibrating model parameters that capture these effects. 

\section{Conclusions}
\label{sec:conclusion}

In this work, we have investigated the dependence of the halo shape and halo formation proxies using the gravity-only MDPL2 N-body simulation. Our main findings are summarized as follows:
\begin{itemize}
    \item In keeping with earlier work, we find that halo ellipticity shows strong dependence on halo formation history, exhibiting the strongest correlations with halo formation proxies such as halo concentration, virial ratio, and peak-centroid offset. Halo ellipticity is additionally correlated with the lookback time to the redshift at which the halo attained half of its present-day mass. 
    \item Halo concentration and peak-centroid offset shows a nearly universal relation with halo ellipticity that is only weakly dependent on halo peak height. 
    \item We have developed and calibrated a probabilistic model for the dependence of halo shape on mass, redshift, and formation history, using Conditional Abundance Matching to capture multi-dimensional correlations.
    \item We presented a Monte Carlo integration technique for modeling halo surface mass density profiles, $\Delta\Sigma(R),$ with capability to incorporate distributions of halo ellipticity and orientation in a straightforward manner. We find that the scatter in $\Delta\Sigma$ carries a signature that may be used to break the degeneracy between halo ellipticity and orientation bias using a forward modeling approach enabled by our model.
\end{itemize}

\section*{Acknowledgements}

We thank the anonymous referee and Joop Schaye for useful comments and suggestions to the paper. 
The CosmoSim database used in this paper is a service by the Leibniz-Institute for Astrophysics Potsdam.  The MultiDark database was developed in cooperation with the Spanish MultiDark Consolider Project CSD2009-00064.
The authors acknowledge the Gauss Centre for Supercomputing e.V. and the Partnership for Advanced Supercomputing in Europe for funding the MultiDark simulation project by providing computing time on the GCS Supercomputer SuperMUC at Leibniz Supercomputing Centre. The simulations have been performed within the Bolshoi project of the University of California High-Performance AstroComputing Center (UC-HiPACC) and were run at the NASA Ames Research Center. 

EL and NC acknowledge the University of Miami for funding this project. DN acknowledges support by National Science Foundation grant AST-1412768. Work done by APH was supported by the U.S. Department of Energy, Office of Science, Office of Nuclear Physics, under contract DE-AC02-06CH11357. We gratefully acknowledge use of the Bebop cluster in the Laboratory Computing Resource Center at Argonne National Laboratory and the facilities and staff of the Yale Center for Research Computing. 

{\em Software}: matplotlib \citep{Hunter:2007}, numpy \citep{van2011numpy}, scipy \citep{ 2020SciPy-NMeth}, astropy \citep{astropy:2013, astropy:2018}, halotools \citep{hearin_etal17}, colossus \citep{colossus}.

\section*{Data Availability}
The data underlying this article are publicly available through the MultiDark Database \citep{riebe_etal13}, and can be downloaded from {\url{https://www.cosmosim.org}}. The scripts used to produce the plots in article are available upon request to the corresponding author.

\bibliographystyle{mnras}
\bibliography{references}

\bsp 
\label{lastpage}
\end{document}